\documentclass[11pt,a4paper]{article}

\usepackage[margin=2.5cm]{geometry}
\usepackage{amsmath,amssymb,amsthm,mathtools}
\usepackage{microtype}
\usepackage{enumitem}
\usepackage{xcolor}
\usepackage{hyperref}
\usepackage[capitalise,nameinlink]{cleveref}

\usepackage{braket}

\hypersetup{
  colorlinks=true,
  linkcolor=blue!55!black,
  urlcolor=blue!55!black,
  citecolor=blue!55!black
}

\numberwithin{equation}{section}
\theoremstyle{plain}
\newtheorem{proposition}{Proposition}[section]
\newtheorem{theorem}[proposition]{Theorem}
\newtheorem{lemma}[proposition]{Lemma}
\newtheorem{corollary}[proposition]{Corollary}
\theoremstyle{remark}
\newtheorem{remark}[proposition]{Remark}

\newcommand{\C}{\mathbb{C}}
\newcommand{\R}{\mathbb{R}}

\newcommand{\ii}{\mathrm{i}}
\newcommand{\ee}{\mathrm{e}}
\newcommand{\dd}{\mathrm{d}}
\newcommand{\TV}{\mathrm{TV}}
\newcommand{\LHP}{\C_-}
\newcommand{\Log}{\operatorname{Log}}
\newcommand{\Arg}{\operatorname{Arg}}
\newcommand{\supp}{\operatorname{supp}}
\newcommand{\diag}{\operatorname{diag}}
\newcommand{\Err}{\mathcal{E}}
\newcommand{\Mset}{\mathcal M_{R,\lambda}}
\newcommand{\Sset}{\mathcal S}
\newcommand{\ip}[2]{\langle #1,#2\rangle}
\newcommand{\norm}[1]{\left\|#1\right\|}
\newcommand{\polylog}{\operatorname{polylog}}
\renewcommand{\vec}[1]{\boldsymbol{#1}}

\title{Optimal Linear Combination of Hamiltonian Simulation for Dissipative Dynamics}
\author{%
  Hongkang Ni\thanks{hongkang@stanford.edu}\qquad
  Lexing Ying\thanks{lexing@stanford.edu}\\[0.5em]
  \small Stanford University, Stanford, CA 94305, United States
}
\date{}

\begin{document}
\maketitle

\begin{abstract}
Linear combination of Hamiltonian simulation (LCHS) is an efficient method for simulating dissipative dynamics on quantum computers. The choice of the LCHS kernel controls the cost of the algorithm. We prove that the kernel family proposed by Ni and Ying~\cite{NY26} minimizes the LCHS query cost in the minimax sense.
For target accuracy $\epsilon$, the minimum cost is asymptotically $\frac{2\ee}{\pi}(\ln\frac1\epsilon-\ln\ln\frac1\epsilon+o(1))$, with optimal leading constant $\frac{2\ee}{\pi}$.
Our lower-bound proof combines weak duality with a realization lemma that approximates extremal dual functions by matrix elements of dissipative evolutions.
\end{abstract}

\section{Introduction}
Solving differential equations is a central task in scientific computing. In this work, we consider solving the ODE
\begin{equation}\label{eq: time dependent ode}
  \frac{\dd}{\dd t}\vec{x}(t) = -A(t) \vec{x}(t)
\end{equation}
on quantum computers. Under the dissipativity condition $A+A^\dagger\succeq0$, linear combination of Hamiltonian simulation (LCHS)~\cite{ALL23,ACL23,LS25,NY26} provides an efficient approach, with optimal initial-state preparation cost and improved dependence on the target precision.

For clarity, we develop the theory primarily for time-independent $A$; the time-dependent extension is discussed below and in~\cite{ALL23,ACL23,LS25,NY26}. Write $A=X+\ii Y$ for the Cartesian decomposition, where $X$ and $Y$ are Hermitian. Dissipativity is equivalent to $X\succeq0$.
LCHS formulas express the propagator $\ee^{-tA}$ of the dissipative
dynamics $\dot{\vec{x}}=-A\vec{x}$ as
\begin{equation}\label{eq:lchs}
  \ee^{-tA}=\int_\R\ee^{-\ii t(kX+Y)}\,\dd\mu(k).
\end{equation}
Each $\ee^{-\ii t(kX+Y)}$ is a unitary evolution implemented by Hamiltonian simulation, and the integral is
implemented using a linear combination of unitaries~\cite{CW12}.
The simplest instance is the Cauchy kernel
$\dd\mu(k)=\rho(k)\,\dd k$ with $\rho(k)=\frac{1}{\pi(1+k^2)}$.

For quantum implementation, the measure $\mu$ is supported in a window $[-R,R]$, and
the relevant parameters are the window size $R$ and the total variation
$\lambda=\|\mu\|_\TV$.
The norm of the largest Hamiltonian that must be simulated grows like $R$, and the
linear-combination-of-unitaries overhead grows like $\lambda$.
A standard figure of merit~\cite{HA25,LS25,NY26} is therefore the weighted cost
$$\Lambda := \lambda R,$$
which we call the \emph{LCHS query cost}.
Once $\mu$ is truncated, \eqref{eq:lchs} holds only approximately, and one
must balance $R$ and $\lambda$ against the approximation error.

Ni and Ying~\cite{NY26} introduced the kernel family $\mu(k) = w_{R,C}(k)\dd k$ with
\begin{equation}\label{eq:ny-kernel}
  w_{R,C}(k)=\frac{\ee^{2C\arctan(1/R)}}{\pi(1+k^2)}
  \exp\Bigl(\ii C\log\frac{R-k}{R+k}\Bigr),\qquad |k|<R,
\end{equation}
where $C\ge0$ is a parameter and we have absorbed the Fourier normalization into the density.
They bounded the error using the local Wiener norm on the half-line and showed
that this family achieves error $\epsilon$ at cost
$\lambda R\sim\frac{2\ee}{\pi}\ln\frac1\epsilon$
\cite[Theorem~22]{NY26}.
This improved the best previously known constant by a factor of about $2.1$.
We prove that their kernel family is exactly optimal in the uniform
worst-case model.

\subsection{Main results}
For a finite complex Borel measure $\mu$ on $\R$, define the worst-case LCHS error by
\begin{equation}\label{eq:matrix-error}
  \Err(\mu)=
  \sup_{d\ge1,\ X\succeq0,\ Y=Y^*}
  \norm{\ee^{-(X+\ii Y)}-\int\ee^{-\ii(kX+Y)}\,\dd\mu(k)},
\end{equation}
where $X,Y$ range over $d\times d$ complex matrices and $\|\cdot\|$ is the
operator norm.
The matrices are finite-dimensional, with no bound on their dimension
or norms. We omit $t$ in \eqref{eq:lchs} because it can be absorbed into $X$ and $Y$.
We write $\supp\mu$ for the support of the measure $\mu$, and
\[
  \Mset=\bigl\{\mu:\ \supp\mu\subset[-R,R],\ \|\mu\|_\TV\le\lambda\bigr\}.
\]
For $R>0$ let
\begin{equation}\label{eq:pq}
  p=p_R=\frac2\pi\arctan R,\qquad
  q=q_R=1-p=\frac2\pi\arctan\frac1R .
\end{equation}

\begin{theorem}\label{thm:main}
For every $R>0$ and $\lambda\ge0$,
\begin{equation}\label{eq:main}
  \min_{\mu\in\Mset}\Err(\mu)
  =V_R(\lambda):=
  \begin{cases}
    1-\lambda, & 0\le\lambda\le p,\\[2mm]
    q\left(\dfrac p\lambda\right)^{p/q}, & \lambda\ge p.
  \end{cases}
\end{equation}
\end{theorem}
The theorem determines the smallest achievable LCHS error for any budget $(R, \lambda)$. Optimizing the LCHS query cost $\Lambda = \lambda R$ yields the following corollary.

\begin{corollary}\label{cor:cost}
For $0<\epsilon<1$, let
\begin{equation}\label{eq:cost}
  J(\epsilon)=\inf\bigl\{\lambda R :\ R>0,\ \supp\mu\subset[-R,R],\ \|\mu\|_\TV\le\lambda,\
  \Err(\mu)\le\epsilon\bigr\}
\end{equation}
be the smallest cost $\lambda R$ that achieves worst-case error $\epsilon$.
Then, as $\epsilon\to0$,
\begin{equation}\label{eq:cost-asymptotics}
  J(\epsilon)=\frac{2\ee}{\pi}\Bigl(\ln\frac1\epsilon-\ln\ln\frac1\epsilon\Bigr)+o(1).
\end{equation}
\end{corollary}

The upper bound ``$\le$'' in \eqref{eq:main} is essentially due to Ni and
Ying~\cite[Section 5]{NY26}, who use the kernel \eqref{eq:ny-kernel}.
The asymptotic upper bound in \eqref{eq:cost-asymptotics} follows from the same construction after optimizing its parameters.
Our contribution is to establish matching lower bounds.

\begin{remark}[Time-dependent generators]\label{rem:time-dependent}
\Cref{thm:main} and \Cref{cor:cost} remain valid for
time-dependent generators $A(s)=X(s)+\ii Y(s)$.
For a measure $\mu$ define
\begin{equation}\label{eq:matrix-error-td}
  \widetilde\Err(\mu)=
  \sup\norm{\mathcal T\ee^{-\int_0^t(X(s)+\ii Y(s))\,\dd s}
  -\int\mathcal T\ee^{-\ii\int_0^t(kX(s)+Y(s))\,\dd s}\,\dd\mu(k)},
\end{equation}
where the supremum is over $d\ge1$, $t>0$, and integrable families of
$d\times d$ matrices with $X(s)\succeq0$ and $Y(s)=Y(s)^*$ for
$s\in[0,t]$.
Then 
\begin{equation}\label{eq: ter eq vr}
\min_{\mu\in\Mset}\widetilde\Err(\mu)=V_R(\lambda).
\end{equation}
The lower bound follows immediately from \Cref{thm:main} because
$\widetilde\Err(\mu)\ge\Err(\mu)$ by definition. In fact, the lower bound in \eqref{eq: ter eq vr} holds even when the matrix dimension is restricted to $d=2$, as discussed in \Cref{rem:td-realization}.
The upper bound follows because the estimate used in
\Cref{sec:upper} is also valid for time-ordered evolutions
\cite[Theorem~20]{NY26}.
\end{remark}

\subsection{Method}
The lower bound comes from weak duality.
For unit vectors $\vec{u},\vec{v}$, consider the matrix-element function
$\varphi(z)=\ip{\vec{v}}{\ee^{-\ii(zX+Y)}\vec{u}}$.
Since $\|\Delta\|=\sup_{\vec{u},\vec{v}}|\ip{\vec{v}}{\Delta \vec{u}}|$ for every matrix $\Delta$,
the worst-case error is a supremum of absolute values of
affine functionals of $\mu$:
\[
  \Err(\mu)=\sup_{\varphi}\Bigl|\varphi(-\ii)-\int\varphi\,\dd\mu\Bigr| .
\]
Each such $\varphi$ is a bounded analytic function on the lower half-plane
$\LHP=\{\operatorname{Im}z<0\}$, and each gives the lower bound
$\Err(\mu)\ge|\varphi(-\ii)|-\lambda\max_{[-R,R]}|\varphi|$.
If we optimize this bound over \emph{all} analytic functions bounded by one, a
harmonic measure argument produces the value $V_R(\lambda)$ and an explicit
extremal family $\Phi_{R,a}$.
The remaining question is whether matrix elements can approximate these
extremal functions.
Our \emph{realization lemma} shows that they can: every bounded analytic
function on $\LHP$ that extends analytically across a compact set
$K\subset\R$ is, up to arbitrarily small error on $K\cup\{-\ii\}$, a matrix
element $\ip{\vec{v}}{\ee^{-\ii(zX+Y)}\vec{u}}$ with $X\succeq0$ and $Y=Y^*$.
Its proof combines polynomial approximation, generalized quantum signal
processing~\cite{MW24}, and an autonomous clock construction.

\subsection{Related work}
LCHS was introduced by An, Liu and Lin~\cite{ALL23} with the Cauchy kernel,
which achieves optimal state preparation cost but has only polynomial decay,
so that truncation to $[-R,R]$ requires $R=O(1/\epsilon)$.
An, Childs and Lin~\cite{ACL23} and Huang and An~\cite{HA25} replaced it by kernels with
near-exponential decay, improving the precision dependence to
$O(\polylog\frac1\epsilon)$.
Low and Somma~\cite{LS25} obtained a kernel with cost
$O(\log\frac1\epsilon)$ through a contour-integral construction.
Ni and Ying~\cite{NY26} reduced LCHS error analysis to scalar Fourier
approximation problems via Weyl calculus.
Their kernel~\eqref{eq:ny-kernel} improves the asymptotic constant of
\cite{LS25} by a factor of about $2.1$, and our results show that this constant
cannot be improved in the worst-case model.
LCHS-type formulas have also been developed for more general matrix
functions~\cite{ACLY26,NY26}.

Most previous work on LCHS concerns upper bounds: constructions of kernels with small cost.
Low and Somma~\cite{LS25} also studied a related bounded polynomial approximation problem numerically. To our knowledge, the present paper gives the first sharp lower bound for
the cost of LCHS that holds uniformly over all matrices.

While preparing this manuscript, we became aware of concurrent work~\cite{AB26} 
on LCHS cost lower bounds based on Hardy space theory. Both works obtain the same asymptotic lower bound, but our result applies to a larger class of kernels and directly bounds the operator-norm residual. Specifically, our lower bound allows arbitrary finite complex measures, whereas~\cite{AB26} requires the admissible kernels to have a prescribed pole and satisfy Hardy-space conditions. Our lower bound concerns the LCHS residual itself, whereas that of~\cite{AB26} concerns an upper bound on the residual used in~\cite{LS25}.

\subsection{Organization of the paper}
\Cref{sec:duality} establishes weak duality, introduces the dual
functions $\Phi_{R,a}$, states the realization lemma, and derives the lower
bound in \Cref{thm:main} from it.
\Cref{sec:realization} proves the realization lemma, whose construction ultimately uses only
finite-dimensional matrices.
\Cref{sec:upper} recalls the upper bound of~\cite{NY26} and completes
the proof of \Cref{thm:main}.
\Cref{sec:cost} proves \Cref{cor:cost}.
\Cref{app:technical} collects two elementary lemmas used in
\Cref{sec:realization}.

\subsection{AI usage and formal verification}

The authors use AI models, particularly ChatGPT 6 Astra and
Claude Opus 5.5, for assistance in exploring proof strategies, performing calculations, and developing the Lean~4 formalization. The authors take responsibility for verifying the results and for their exposition.

\Cref{thm:main} has been formalized and checked in the Lean~4 proof assistant~\cite{Lean4} with the Mathlib library~\cite{Mathlib}. The formalization is available in the
\href{https://github.com/HongkangNi/Optimal-LCHS-for-Dissipative-Dynamics}
{companion GitHub repository}. The formal proof depends only on Lean's standard axioms. \Cref{cor:cost} and the time-dependent extension in \Cref{rem:time-dependent} are not part of the formalization. The formal proof derives the upper bound independently from the Poisson formula on the lower half-plane, and does not rely on~\cite{NY26}.

\section{Weak duality and the lower bound}\label{sec:duality}

\subsection{Weak duality}\label{sec:weak}

We use the inner product $\ip{\vec{u}}{\vec{v}}=\sum_j\overline{u_j}v_j$, which is linear in the
second argument.

Let the set of matrix-element functions be
\begin{equation}\label{eq:S}
  \Sset=\Bigl\{\,\varphi(z) = \ip{\vec{v}}{\ee^{-\ii(zX+Y)}\vec{u}}\ :\
  \begin{aligned}
  &d\ge1,\ X,Y\in\C^{d\times d},\ X\succeq0,\ Y=Y^*,\\
  &\vec{u},\vec{v}\in\C^d,\ \|\vec{u}\|=\|\vec{v}\|=1
  \end{aligned}\Bigr\}.
\end{equation}
Every $\varphi\in\Sset$ is entire and, by \Cref{lem:contraction}, satisfies
$|\varphi|\le1$ on $\overline{\LHP}$.
In particular, $\varphi(-\ii)=\ip{\vec{v}}{\ee^{-(X+\ii Y)}\vec{u}}$.

Since $\|\Delta\|=\sup_{\|\vec{u}\|=\|\vec{v}\|=1}|\ip{\vec{v}}{\Delta \vec{u}}|$ for every matrix
$\Delta$ and the inner product can be
taken inside the integral,
\begin{equation}\label{eq:E-sup}
  \Err(\mu)=\sup_{\varphi\in\Sset}\Bigl|\varphi(-\ii)-\int\varphi\,\dd\mu\Bigr| .
\end{equation}
Thus, the worst-case error is the supremum of the absolute values of affine
functionals of $\mu$, indexed by $\Sset$.
Exchanging the infimum over $\mu$ and the supremum over $\varphi$ gives the
weak duality bound
\begin{equation}\label{eq:weak-duality}
\begin{aligned}
  \inf_{\mu\in\Mset}\Err(\mu)
  &=\inf_{\mu\in\Mset}\ \sup_{\varphi\in\Sset}
    \Bigl|\varphi(-\ii)-\int\varphi\,\dd\mu\Bigr|\\
  &\ge\sup_{\varphi\in\Sset}\ \inf_{\mu\in\Mset}
    \Bigl|\varphi(-\ii)-\int\varphi\,\dd\mu\Bigr|\\
  &\ge\sup_{\varphi\in\Sset}
    \Bigl[\,|\varphi(-\ii)|-\lambda\max_{[-R,R]}|\varphi|\Bigr].
\end{aligned}
\end{equation}
The last step follows from the reverse triangle inequality and
$|\int\varphi\,\dd\mu|\le\|\mu\|_\TV\max_{[-R,R]}|\varphi|$ for
$\mu\in\Mset$.
Every $\varphi\in\Sset$ therefore gives a lower bound on the error of every
admissible $\mu$.
It remains to choose $\varphi$.

\begin{remark}[Strong duality]
Replacing $|\cdot|$ by $\operatorname{Re}(\cdot)$ in \eqref{eq:E-sup} does
not change the supremum, because $\Sset$ is invariant under
$\vec{u}\mapsto\ee^{\ii\theta}\vec{u}$. The resulting expression satisfies strong duality: Sion's minimax theorem shows that equality holds throughout
\eqref{eq:weak-duality}. We do not need this fact, since the lower bound uses only weak duality and the upper bound in \Cref{sec:upper} follows from an explicit kernel construction.
\end{remark}

\subsection{The relaxed dual problem}\label{sec:relaxed}

Since every $\varphi\in\Sset$ is analytic and bounded by $1$ on $\LHP$, it is
natural to relax \eqref{eq:weak-duality} by allowing all such functions. This gives the relaxed problem
\begin{equation}\label{eq:relaxed}
  \sup_{\varphi\in\overline{\Sset}}
    \Bigl[\,|\varphi(-\ii)|-\lambda\max_{[-R,R]}|\varphi|\Bigr],
\end{equation}
where
\[
  \overline{\Sset}:= \{\varphi(z): \varphi(z)\text{ analytic on } \LHP \text{ and continuous on }\overline{\LHP},\ |\varphi(z)|\le 1\text{ on } \LHP\}.
\]
Although the lower-bound proof in \Cref{sec:lower} does not depend on solving this relaxed problem, the latter motivates our construction of the dual functions.

Let $\Phi\in\overline{\Sset}$, and let $a=\max_{[-R,R]}|\Phi|\in(0,1]$.
The function $\log|\Phi|$ is subharmonic and bounded above.
On the boundary, it is at most $\log a$ on $[-R,R]$ and at most $0$ on the
rest of $\R$.
The two-constants theorem~\cite[Theorem~4.3.7]{Ran95} gives
\begin{equation}\label{eq: two-const}
  \log|\Phi(-\ii)|\le\omega\log a,
\end{equation}
where $\omega$ is the harmonic measure of $[-R,R]$ viewed from $-\ii$.
It is given by integrating the Cauchy density:
\[
  \omega=\int_{-R}^R\rho(k)\,\dd k=\frac2\pi\arctan R=p .
\]
Hence,
\begin{equation}\label{eq:relaxed scalar prob}
|\Phi(-\ii)|-\lambda\max_{[-R,R]}|\Phi|\le a^p-\lambda a.
\end{equation}
The formal extremal for the bound $|\Phi(-\ii)|\le a^p$ is the outer function with modulus $a$ on $(-R,R)$ and $1$ outside.
A more elementary way to obtain this formal extremal is to require equality in \eqref{eq: two-const}, for which $\log|\Phi|$ must be harmonic. Thus, $\log|\Phi|$ is the bounded solution of the Dirichlet problem for the Laplace equation in the lower half-plane with boundary conditions
$$\log|\Phi(x)| = \log a\ (x\in(-R,R)),\qquad \log|\Phi(x)| = 0\ (x\in\mathbb{R}\backslash[-R,R]).$$
This determines $\Phi$ up to a unimodular constant and yields the function $\Phi_{R,a}$ defined in \eqref{eq:dual-function} below. For $a<1$, this function has no continuous boundary values at $\pm R$ and therefore does not belong to $\overline{\Sset}$; we use regularized versions in the rigorous proof.

This motivates maximizing $a^p-\lambda a$ over
$a\in(0,1]$, which gives the candidate value $V_R(\lambda)$.
Since $\Sset$ is contained in the relaxed set, weak duality can never give a
lower bound larger than $V_R(\lambda)$.
The realization lemma (\Cref{lemma:realization}) allows us to approximate
$\Phi_{r,a}$ on $[-R,R]\cup\{-\ii\}$ for $r>R$ by matrix elements.
Letting $r\downarrow R$ then gives the matching lower bound, as shown in \Cref{sec:lower}.

\subsection{The dual functions}\label{sec:dual functions}
We now define the dual functions used in the lower-bound proof.
Given $0<a\le1$, write $a=\ee^{-\pi c}$ with $c\ge0$.
Let $M_R(z)=\frac{R-z}{R+z}$.
Since $\operatorname{Im}M_R(z)=-2R\operatorname{Im}z/|R+z|^2$, the map $M_R$
sends $\LHP$ into the upper half-plane.
Define
\begin{equation}\label{eq:dual-function}
  \Phi_{R,a}(z)=\ee^{-\pi c}
  \exp\Bigl[-\ii c\Log\frac{R-z}{R+z}\Bigr],\qquad z\in\LHP,
\end{equation}
with the branch of the logarithm for which
$0<\Arg\frac{R-z}{R+z}<\pi$. The following lemma establishes the required properties.

\begin{lemma}\label{lem:dual-function}
$\Phi_{R,a}$ is holomorphic on $\LHP$.
It extends continuously to $\overline{\LHP}\setminus\{\pm R\}$, and
holomorphically across $(-R,R)$.
Moreover,
\begin{equation}\label{eq:dual-properties}
  \sup_{\LHP}|\Phi_{R,a}|\le1,\qquad
  |\Phi_{R,a}(k)|=
  \begin{cases}a,&k\in (-R,R),\\ 1,&k\in \mathbb{R}\backslash[-R,R],\end{cases}\qquad
  \Phi_{R,a}(-\ii)=a^p .
\end{equation}
\end{lemma}

\begin{proof}
The modulus is $|\Phi_{R,a}(z)|=\exp\bigl(-c(\pi-\Arg M_R(z))\bigr)\le1$.
On the real axis, $\Arg M_R(k)=0$ for $|k|<R$ and $\Arg M_R(k)=\pi$ for
$|k|>R$.
Finally, $M_R(-\ii)=\frac{R+\ii}{R-\ii}=\ee^{2\ii\arctan(1/R)}=\ee^{\ii\pi q}$.
Hence $\Phi_{R,a}(-\ii)=\ee^{-\pi c}\ee^{\pi cq}=\ee^{-\pi cp}=a^p$.
\end{proof}

\subsection{The realization lemma and the lower bound}\label{sec:lower}

In \Cref{sec:relaxed,sec:dual functions}, we motivated the dual function construction through the relaxed problem \eqref{eq:relaxed}, obtained by enlarging the class of dual functions from $\Sset$ to $\overline{\Sset}$. The following realization lemma shows that this relaxation does not weaken our lower bound: suitable functions in $\overline{\Sset}$ can be approximated by functions in $\Sset$ on the compact sets needed in the proof.

\begin{lemma}[Realization lemma]\label{lemma:realization}
Let $K\subset\R$ be compact.
Let $\Phi$ be holomorphic on a neighbourhood of $\LHP\cup K$, with
$|\Phi|\le1$ on $\LHP$.
Then for every $\delta>0$ there are $d\ge1$, matrices $X\succeq0$ and
$Y=Y^*$ in $\C^{d\times d}$, and unit vectors $\vec{u},\vec{v}\in\C^d$ such that
\begin{equation}\label{eq:realization}
  \sup_{k\in K}\bigl|\ip{\vec{v}}{\ee^{-\ii(kX+Y)}\vec{u}}-\Phi(k)\bigr|<\delta,
  \qquad
  \bigl|\ip{\vec{v}}{\ee^{-(X+\ii Y)}\vec{u}}-\Phi(-\ii)\bigr|<\delta .
\end{equation}
The dimension and the norms of $X$ and $Y$ may depend on $\delta$.
\end{lemma}

We defer the proof to \Cref{sec:realization}. The realization lemma immediately yields the lower bound in \Cref{thm:main}.

\begin{proposition}[Lower bound]\label{prop:lower}
Every $\mu\in\Mset$ satisfies $\Err(\mu)\ge a^p-a\lambda$ for all
$a\in(0,1]$, and hence $\Err(\mu)\ge V_R(\lambda)$.
\end{proposition}

\begin{proof}
Fix $a\in(0,1]$, $r>R$ and $\delta>0$.
Apply \Cref{lemma:realization} with $K=[-R,R]$ and $\Phi=\Phi_{r,a}$.
This gives $\varphi\in\Sset$ with $|\varphi-\Phi_{r,a}|<\delta$ on
$[-R,R]\cup\{-\ii\}$.
By \Cref{lem:dual-function} with $r$ in place of $R$, we have
$\Phi_{r,a}(-\ii)=a^{p_r}$ and $|\Phi_{r,a}|=a$ on $[-R,R]\subset(-r,r)$.
Hence $|\varphi(-\ii)|\ge a^{p_r}-\delta$ and
$\max_{[-R,R]}|\varphi|\le a+\delta$.
By \eqref{eq:weak-duality},
\[
  \Err(\mu)\ \ge\ |\varphi(-\ii)|-\lambda\max_{[-R,R]}|\varphi|
  \ \ge\ a^{p_r}-\lambda a-(1+\lambda)\delta .
\]
Letting $\delta\downarrow0$ and then $r\downarrow R$, so that $p_r\to p$,
gives $\Err(\mu)\ge a^p-\lambda a$.

It remains to maximize $a^p-\lambda a$ over $a\in(0,1]$.
If $\lambda\le p$, take $a=1$ to get $1-\lambda$.
If $\lambda\ge p$, take $a_*=(p/\lambda)^{1/q}\le1$.
At this point, $a_*^{p-1}=\lambda/p$, and the value is
$a_*(\lambda/p-\lambda)=q(p/\lambda)^{p/q}$.
In both cases, the value is $V_R(\lambda)$.
\end{proof}

\section{Proof of the realization lemma}\label{sec:realization}

Let $K$ and $\Phi$ be as in \Cref{lemma:realization}.
The set $Z=K\cup\{-\ii\}$ is a compact subset of $\overline{\LHP}$.
Since $\ee^{-\ii(-\ii X+Y)}=\ee^{-(X+\ii Y)}$, we must approximate $\Phi$
uniformly on $Z$ by matrix elements $\ip{\vec{v}}{\ee^{-\ii(zX+Y)}\vec{u}}$, where
$X\succeq0$ and $Y=Y^*$ are matrices and $\vec{u},\vec{v}$ are unit vectors.
We proceed in four steps, each introducing an arbitrarily small error
uniformly on $Z$:
\begin{enumerate}[leftmargin=2em]
\item approximate $\Phi$ by a polynomial $P$ in $\ee^{-\ii hz}$;
\item write $P(\ee^{-\ii hz})$ as a matrix element of a time-ordered
  evolution generated by $zL(s)+H(s)$ with $L(s)\succeq0$;
\item remove the time dependence with a clock register, which gives
  time-independent but infinite-dimensional operators;
\item discretize the clock, which gives finite matrices.
\end{enumerate}
These steps are carried out in
\Cref{sec:poly,sec:pulses,sec:clock,sec:discretization}, respectively.

\subsection{Approximation by bounded analytic polynomials}\label{sec:poly}

For small $h>0$, define
\[
  F_h(w)=\Phi\Bigl(-\frac{2\ii}{h}\,\frac{1-w}{1+w}\Bigr),\qquad |w|<1 .
\]
The map $w\mapsto-\frac{2\ii}{h}\frac{1-w}{1+w}$ sends the unit disc onto
$\LHP$, so $F_h$ is a Schur function on the disc.
Substituting $w=\ee^{-\ii hz}$ turns the argument of $\Phi$ into
\[
  \frac2h\tan\frac{hz}2\longrightarrow z\qquad(h\to0),
\]
uniformly on bounded sets.
Next, replace $F_h(w)$ by $F_h(\varrho w)$ with $\varrho<1$ close to $1$.
Then truncate its Taylor series and divide by a factor slightly larger than
$1$.
This produces a polynomial $P$ with $|P|\le1$ on the closed unit disc and
\[
  P(\ee^{-\ii hz})\approx\Phi(z)\qquad\text{uniformly for }z\in K\cup\{-\ii\}.
\]
The analyticity of $\Phi$ near $K$ guarantees that the approximation is
uniform up to the boundary.

\subsection{Realization as a time-dependent evolution}\label{sec:pulses}

Replacing $P$ by $(1-\eta)P$ with a small $\eta>0$ changes $P(\ee^{-\ii hz})$
by at most $\eta$ on $Z$, since $|\ee^{-\ii hz}|=\ee^{h\operatorname{Im}z}\le1$
there.
Thus, we may assume that $1-|P|^2>0$ on the unit circle.
Such a polynomial admits a generalized quantum signal processing (GQSP)
decomposition~\cite{MW24}:
\begin{equation}\label{eq:gqsp}
  P(w)=\langle0|\,V_mD(w)V_{m-1}\cdots D(w)V_0\,|0\rangle,
  \qquad D(w)=\diag(w,1),
\end{equation}
with constant unitaries $V_j\in U(2)$.
The decomposition is obtained in two steps.
Fej\'er--Riesz factorization first gives a complementary polynomial $Q$ with
$|P|^2+|Q|^2=1$ on the unit circle.
The unitary polynomial matrix with first column $(P,Q)$ is then split into
layers; see~\cite{MW24,BS25,NSYL25}.

Let $\Pi_0=|0\rangle\langle0|$ and
\begin{equation}\label{eq:Uz def}
  \mathcal U(z)=V_m\,D(\ee^{-\ii hz})\,V_{m-1}\cdots D(\ee^{-\ii hz})\,V_0 .
\end{equation}
Each factor is an exponential: $D(\ee^{-\ii hz})=\ee^{-\ii zh\Pi_0}$, and
$V_j=\ee^{-\ii H_j}$ for some Hermitian $H_j$ by the spectral theorem.
We construct a time-dependent evolution by applying the factors successively, each over a unit time
interval.
Let $\chi(s)=1-\cos(2\pi s)$ for $s\in[0,1]$ and $\chi(s)=0$ otherwise, so
that $\chi\ge0$ and $\int\chi=1$.
With $T_0=2m+2$, define
\begin{equation}\label{eq:L H def}
  H(s)=\sum_{j=0}^m\chi(s-1-2j)\,H_j,\qquad
  L(s)=h\sum_{j=1}^m\chi(s-2j)\,\Pi_0 .
\end{equation}
These functions are continuous, with $L(s)\succeq0$ and $H(s)=H(s)^*$, and both vanish outside
$[1,T_0]$.
The pulse producing $V_j$ occupies $[1+2j,2+2j]$, and the pulse producing
the $j$-th factor $D(\ee^{-\ii hz})$ occupies $[2j,2j+1]$.
Let $U_z(s)$ be the solution of
\begin{equation}\label{eq:ds Uz}
  \frac{\dd}{\dd s}U_z(s)=-\ii\bigl(zL(s)+H(s)\bigr)U_z(s),\qquad U_z(0)=I,
\end{equation}
that is,
$$U_z(s)=\mathcal T e^{-\ii\int_0^s(zL(r)+H(r))\,\dd r}.$$ 
Since the supports of $H(s)$ and $L(s)$ are disjoint and alternate, the evolution multiplies the factors in \eqref{eq:Uz def} in the required order. Therefore,
\begin{equation}\label{eq:pulses}
  U_z(s)=I\quad(s\le1),\qquad U_z(s)=\mathcal U(z)\quad(s\ge T_0).
\end{equation}
In particular,
\begin{equation}\label{eq:time-ordered}
  P(\ee^{-\ii hz})=\langle0|\,\mathcal U(z)\,|0\rangle
  =\langle0|\,\mathcal T\ee^{-\ii\int_0^{T_0}(zL(s)+H(s))\,\dd s}\,|0\rangle .
\end{equation}

\begin{remark}[A time-dependent realization lemma]\label{rem:td-realization}
The construction above already proves a time-dependent version of
\Cref{lemma:realization}: $\Phi$ is approximated on $Z$ by the
time-ordered matrix element
$$\phi(z)=\langle0|\,\mathcal T\ee^{-\ii\int_0^{T_0}(zL(s)+H(s))\,\dd s}\,|0\rangle,$$
with $L(s)\succeq0$ and $H(s)=H(s)^*$.
Here $L(s)$ and $H(s)$ are $2\times2$ matrices, and the vectors are
$\vec{u}=\vec{v}=|0\rangle$, so this time-dependent realization uses only
dimension $d=2$.
The weak duality bound \eqref{eq:weak-duality} also holds for the
time-dependent error $\widetilde\Err$ of \eqref{eq:matrix-error-td}, with
$\Sset$ replaced by the corresponding set of time-ordered matrix elements.
Applying the proof of \Cref{prop:lower} to this version
therefore gives $\widetilde\Err(\mu)\ge V_R(\lambda)$ directly, without the
clock construction below.
The clock is needed only for the stronger, time-independent statement
$\Err(\mu)\ge V_R(\lambda)$.
\end{remark}

\subsection{An autonomous clock}\label{sec:clock}

To remove the time dependence, we introduce a clock register on which a wave packet moves at unit speed, so that the internal state encounters the pulses $L(s)$, $H(s)$ in sequence.
Let $\ell\ge T_0+2$ be an integer.
We regard $L$ and $H$ as $\ell$-periodic functions, which is consistent
because they vanish on $[0,1]$ and on $[T_0,\ell]$.
On the clock space $L^2(\R/\ell\R)\otimes\C^2$, define operators
\[
  \mathsf L=L(s)\succeq0,\qquad\mathsf H=-\ii\partial_s+H(s)=\mathsf H^* .
\]

Let $g(s)=\sqrt{128/35}\,\sin^4(\pi s)$ on $[0,1]$ and $g=0$ elsewhere.
This is a real $C^2$ packet supported in $[0,1]$ with $\int g^2=1$.
Let $g_{\rm end}=g(\cdot-T_0)$ denote the same packet translated by $T_0$.

The following lemma shows that the operators $T_0\mathsf L$ and $T_0\mathsf H$ give the desired realization, although they act on an infinite-dimensional space and the latter is unbounded. We discretize these operators to obtain finite-dimensional matrices in \Cref{sec:discretization}.

\begin{lemma}[Continuous clock]\label{lem:continuous-clock}
Let
\[
  \vec{u}=g\otimes|0\rangle,\qquad \vec{v}=g_{\rm end}\otimes|0\rangle ,
\]
which are unit vectors in $L^2(\R/\ell\R)\otimes\C^2$.
For $z\in Z$,
\begin{equation}\label{eq:continuous-clock}
  \ip{\vec{v}}{\ee^{-\ii T_0(z\mathsf L+\mathsf H)}\vec{u}}
  =\langle0|\,\mathcal T\ee^{-\ii\int_0^{T_0}(zL(s)+H(s))\,\dd s}\,|0\rangle
  =P(\ee^{-\ii hz}).
\end{equation}
\end{lemma}

\begin{proof}
The evolution $\ee^{-\ii t(z\mathsf L+\mathsf H)}\vec{u}$ solves the initial-value problem
\begin{equation}\label{eq:transport}
  \begin{aligned}
  \ii\,\partial_t\vec{\Psi}_z(t,s)&=-\ii\,\partial_s\vec{\Psi}_z(t,s)
  +\bigl(zL(s)+H(s)\bigr)\vec{\Psi}_z(t,s),\\
  \vec{\Psi}_z(0,s)&=g(s)\,|0\rangle .
  \end{aligned}
\end{equation}
Solving this transport equation by the method of characteristics gives
\begin{equation}\label{eq:reference solution}
  \vec{\Psi}_z(t,s)=g(s-t)\,U_z(s)|0\rangle,\qquad t\in[0,T_0],\ s\in\R .
\end{equation}
Using \eqref{eq:ds Uz}, one can also verify directly that $\vec{\Psi}_z$ satisfies \eqref{eq:transport}.

By \eqref{eq:pulses}, $U_z=I$ on the support of $g$ and $U_z=\mathcal U(z)$
on the support of $g_{\rm end}$, so
\begin{equation}\label{eq:endpoints}
  \vec{\Psi}_z(0,\cdot)=g\,|0\rangle,\qquad
  \vec{\Psi}_z(T_0,\cdot)=g_{\rm end}\,\mathcal U(z)|0\rangle .
\end{equation}
Moreover, $\vec{\Psi}_z(t,\cdot)$ is supported in $[t,t+1]\subset[0,\ell-1]$: the
packet never crosses the point $0\equiv\ell$ of the circle, so
periodicity plays no role.

Since $-\ii(z\mathsf L+\mathsf H)$ is a bounded perturbation of the generator
$-\partial_s$ of the translation group, such solutions are
unique~\cite{Paz83}, and $\ee^{-\ii t(z\mathsf L+\mathsf H)}\vec{u}=\vec{\Psi}_z(t,\cdot)$.
By \eqref{eq:endpoints} and $\|g_{\rm end}\|=1$, we obtain
$\ip{\vec{v}}{\ee^{-\ii T_0(z\mathsf L+\mathsf H)}\vec{u}}=\langle0|\,\mathcal U(z)\,|0\rangle$,
and \eqref{eq:time-ordered} gives \eqref{eq:continuous-clock}.
\end{proof}


\subsection{Discretization}\label{sec:discretization}

We replace the circle $\R/\ell\R$ by a periodic grid with mesh size $1/n$ and
approximate $-\ii\partial_s$ by a central difference.
Let $N=n\ell$ and $s_j=j/n$, with $j$ taken modulo $N$.
On $(\C^2)^N=\C^N\otimes\C^2$ we define the cyclic shift
$(\mathsf T\vec{\psi})_j=\vec{\psi}_{j+1}$, the Hermitian matrices
\[
  \mathsf D=\frac{n}{2\ii}(\mathsf T-\mathsf T^*),\qquad
  \mathsf L_n=\diag\bigl(L(s_j)\bigr)_j\succeq0,\qquad
  \mathsf H_n=\diag\bigl(H(s_j)\bigr)_j,
\]
and the generator $\mathsf G(z)=z\mathsf L_n+\mathsf D+\mathsf H_n$. We also define the sampling map $(\mathsf S f)_j=n^{-1/2}f(s_j)$.
By \Cref{lem:contraction}, $\|\ee^{-\ii t\mathsf G(z)}\|\le1$ for
$t\ge0$ and $\operatorname{Im}z\le0$.

\paragraph{The sampled reference solution is an approximate solution.}
Let $\vec{\psi}_z(t)=\mathsf S\vec{\Psi}_z(t,\cdot)$, where $\vec{\Psi}_z$ is the reference solution in \eqref{eq:reference solution}.
Sampling \eqref{eq:transport} gives
\[
  \ii\,\vec{\psi}_z'(t)=\mathsf G(z)\vec{\psi}_z(t)+\vec{r}_z(t),
\]
with the residual given by
\[
  \vec{r}_z(t)=\mathsf S\bigl(-\ii\partial_s\vec{\Psi}_z(t,\cdot)\bigr)-\mathsf D\vec{\psi}_z(t),
\]
which is the consistency error of the central difference.
Since the packet stays away from the seam, the wrap-around in $\mathsf T$
plays no role. We have
\[
  (\vec{r}_z(t))_j=\frac{\ii\,n^{1/2}}{2}
  \int_{-1/n}^{1/n}\bigl(\partial_s\vec{\Psi}_z(t,s_j+\theta)-\partial_s\vec{\Psi}_z(t,s_j)\bigr)
  \,\dd\theta .
\]
Let
\[
  M=\sup_{z\in Z,\ t\in[0,T_0],\ s\in\R}\norm{\partial_{ss}\vec{\Psi}_z(t,s)}<\infty .
\]
Then the integrand is bounded by $M|\theta|$, so
$\|(\vec{r}_z(t))_j\|\le\frac{M}{2}\,n^{-3/2}$ and
$\|\vec{r}_z(t)\|\le\frac{\sqrt\ell\,M}{2n}$.
The contraction property and the Duhamel estimate (\Cref{lem:duhamel}) give
\begin{equation}\label{eq:clock-duhamel}
  \bigl\|\vec{\psi}_z(T_0)-\ee^{-\ii T_0\mathsf G(z)}\vec{\psi}_z(0)\bigr\|
  \le\frac{T_0\sqrt\ell\,M}{2n}\longrightarrow0
  \qquad(n\to\infty),
\end{equation}
uniformly for $z\in Z$.

\paragraph{Reading off the matrix element.}
By \eqref{eq:endpoints}, $\vec{\psi}_z(0)=(\mathsf Sg)\otimes|0\rangle$, which does
not depend on $z$, and $\vec{\psi}_z(T_0)=(\mathsf Sg_{\rm end})\otimes\mathcal U(z)|0\rangle$.
Since $nT_0$ is an integer, $\mathsf S g_{\rm end}$ is $\mathsf S g$ shifted
by $nT_0$ sites, and their common squared norm satisfies
$\sigma_n^2:=\|\mathsf S g\|^2=\frac1n\sum_jg(s_j)^2\to\int g^2=1$.
Let
\[
  \vec{u}=\frac{1}{\sigma_n}(\mathsf Sg)\otimes|0\rangle,\qquad
  \vec{v}=\frac{1}{\sigma_n}(\mathsf Sg_{\rm end})\otimes|0\rangle,\qquad
  X=T_0\mathsf L_n\succeq0,\qquad Y=T_0(\mathsf D+\mathsf H_n)=Y^* .
\]
Then $\ip{\vec{v}}{\vec{\psi}_z(T_0)}=\sigma_n\langle0|\mathcal U(z)|0\rangle$, and
\eqref{eq:clock-duhamel} gives
\[
  \bigl|\ip{\vec{v}}{\ee^{-\ii(zX+Y)}\vec{u}}-\langle0|\,\mathcal U(z)\,|0\rangle\bigr|
  \le\frac{T_0\sqrt\ell\,M}{2n\,\sigma_n}\longrightarrow0
\]
uniformly for $z\in Z$.

Combining the four steps proves \Cref{lemma:realization}.
The dimension $d=2N$ and the norm $\|\mathsf D\|\sim n$ grow as the approximation
error decreases, as permitted by the lemma.


\section{The upper bound}\label{sec:upper}

The upper bound follows from the results of Ni and Ying~\cite{NY26}.
We recall these results in our normalization.
For a function $f$ on $[0,\infty)$, the \emph{local Wiener norm} is defined by
\[
  \|f\|_{W([0,\infty))}=\inf\Bigl\{\|\nu\|_\TV:\ f(x)=\int\ee^{-\ii kx}\,
  \dd\nu(k)\ \text{for all }x\ge0\Bigr\}.
\]
Their kernel $\hat g_{R,C}$ equals $\sqrt{2\pi}\,w_{R,C}$, and their
normalization factor $\lambda_{\hat g_{R,C}}$ equals
$\int_{-R}^R|w_{R,C}(k)|\,\dd k$.
For a measure $\mu$, we write $f_\mu(x)=\int\ee^{-\ii kx}\,\dd\mu(k)$.

\begin{lemma}[{\cite[Theorem~20]{NY26}}]\label{lem:transfer}
Let $\mu$ be a finite complex measure with compact support.
Then $\Err(\mu)\le\norm{\ee^{-x}-f_\mu}_{W([0,\infty))}$.
\end{lemma}

Theorem~20 of~\cite{NY26} is stated for time-ordered evolutions
$\mathcal T\ee^{-\int_0^tA(s)\,\dd s}$ with $A(s)=L(s)+\ii H(s)$ and
$L(s)\succeq0$.
\Cref{lem:transfer} is the special case $A(s)\equiv X+\ii Y$, $t=1$.
Since $\mu$ has compact support, $f_\mu$ is entire, so the smoothness
hypothesis of that theorem is satisfied.

\begin{lemma}[{\cite[Lemma~21 and the proof of Theorem~22]{NY26}}]
\label{lem:ny-kernel}
Let $C\ge0$ and $\dd\mu=w_{R,C}\mathbf 1_{(-R,R)}\,\dd k$. Then
\begin{align*}
  \|\mu\|_\TV&=\frac2\pi\,\ee^{2C\arctan(1/R)}\arctan R=p\,\ee^{\pi Cq},\\
  \norm{\ee^{-x}-f_\mu}_{W([0,\infty))}
  &\le\frac2\pi\,\ee^{-2C\arctan R}\arctan\frac1R=q\,\ee^{-\pi Cp}.
\end{align*}
\end{lemma}

The alternative expressions follow from $2\arctan R=\pi p$ and
$2\arctan(1/R)=\pi q$.

\begin{proposition}[Upper bound]\label{prop:upper}
For every $\lambda\ge0$, there exists $\mu\in\Mset$ with $\|\mu\|_\TV=\lambda$
and $\Err(\mu)\le V_R(\lambda)$.
\end{proposition}

\begin{proof}
By \Cref{lem:transfer}, it suffices to bound the Wiener error.

\emph{Case $\lambda\ge p$.}
Take $\dd\mu=w_{R,C}\mathbf 1_{(-R,R)}\,\dd k$ with
$C=\log(\lambda/p)/(\pi q)\ge0$.
By \Cref{lem:ny-kernel}, $\|\mu\|_\TV=p\,\ee^{\pi Cq}=\lambda$ and the
Wiener error is at most $q\,\ee^{-\pi Cp}=q(p/\lambda)^{p/q}=V_R(\lambda)$.

\emph{Case $\lambda<p$.}
Let $\alpha=\lambda/p$ and $\dd\mu=\alpha\rho\,\mathbf 1_{(-R,R)}\,\dd k$, so
that $\|\mu\|_\TV=\alpha p=\lambda$.
Since $\int\ee^{-\ii kx}\rho(k)\,\dd k=\ee^{-|x|}$, we have for $x\ge0$
\[
  \ee^{-x}-f_\mu(x)=\int\ee^{-\ii kx}\,\dd\nu(k),\qquad
  \dd\nu=(1-\alpha)\rho\,\dd k+\alpha\rho\,\mathbf 1_{\{|k|>R\}}\,\dd k .
\]
Hence the Wiener error is at most
$\|\nu\|_\TV=(1-\alpha)+\alpha q=1-\alpha p=1-\lambda$.
\end{proof}

\begin{proof}[Proof of \Cref{thm:main}]
By \Cref{prop:lower}, every $\mu\in\Mset$ has
$\Err(\mu)\ge V_R(\lambda)$.
By \Cref{prop:upper}, there exists $\mu\in\Mset$ with
$\Err(\mu)\le V_R(\lambda)$.
\end{proof}

\begin{remark}[A primal--dual optimal pair]
For $\lambda\ge p$, the kernel $w_{R,C}$ and the dual function $\Phi_{R,a}$
with $a=\ee^{-\pi C}$ form a primal--dual optimal pair.
On $(-R,R)$ we have
\[
  w_{R,C}(k)\,\Phi_{R,a}(k)=\ee^{-\pi Cp}\rho(k)>0 .
\]
Thus the phase of the kernel exactly cancels that of $\Phi_{R,a}$, and
$|\Phi_{R,a}|=a$ on the support of $\dd\mu=w_{R,C}\mathbf 1_{(-R,R)}\,\dd k$.
Hence the reverse triangle inequality in \eqref{eq:weak-duality} is an
equality for this pair, and
\[
  \Phi_{R,a}(-\ii)-\int\Phi_{R,a}\,\dd\mu=a^p-p\,\ee^{-\pi Cp}
  =q\,\ee^{-\pi Cp}.
\]
This matches the error bound in \Cref{lem:ny-kernel}.
In particular, the error of the truncated Ni--Ying kernel is exactly
$q\,\ee^{-\pi Cp}=V_R(\lambda)$.
\end{remark}

\section{The optimal cost}\label{sec:cost}

In this section, we prove \Cref{cor:cost}. By \Cref{thm:main}, it suffices to solve a constrained optimization problem in two variables:
\begin{equation}\label{eq:opt prob}
  J(\epsilon) = \min_{\lambda, R} \lambda R\quad \text{s.t.}\quad V_R(\lambda)\le \epsilon.
\end{equation}
The proof is straightforward. We first eliminate the parameter $\lambda$ using the constraint, then differentiate to characterize the optimum, and finally analyze the asymptotics as $\epsilon\to 0$.

\paragraph{Reduction to one parameter.} Set $\beta=\ln(1/\epsilon)$.
By \Cref{thm:main},
$$J(\epsilon)=\inf\{\lambda R:\ R>0,\ \lambda\ge0,\ V_R(\lambda)\le\epsilon\}.$$
We parametrize $R=\cot\frac{\pi q}2$ by $q=q_R\in(0,1)$, with $p=1-q$.
Since $V_R$ is continuous and decreasing in $\lambda$ with $V_R(p)=q$, the
smallest admissible $\lambda$ solves $V_R(\lambda)=\epsilon$.
For $q\ge\epsilon$, this gives $\lambda=p\,(q/\epsilon)^{q/p}$.
For $q<\epsilon$, it gives $\lambda=1-\epsilon$, and the resulting cost
$(1-\epsilon)\cot\frac{\pi q}2$ exceeds the value of the first case at
$q=\epsilon$.
Hence
\begin{equation}\label{eq:J-F}
  J(\epsilon)=\inf_{\epsilon\le q<1}F(q),\qquad
  F(q)=\cot\frac{\pi q}2\,(1-q)\,
  \exp\Bigl[\frac{q}{1-q}\bigl(\beta+\ln q\bigr)\Bigr].
\end{equation}

\paragraph{The optimality equation.}
Let $h(q)=\pi(1-q)^2/\sin(\pi q)$.
A direct computation gives
\[
  \frac{\dd}{\dd q}\ln F(q)=\frac{\beta+\ln q-h(q)}{(1-q)^2}.
\]
The function $h$ decreases from $+\infty$ to $0$ on $(0,1)$, so the
numerator is strictly increasing.
It is negative at $q=\epsilon$, where $\beta+\ln q=0$, and tends to
$\beta>0$ as $q\to1$.
Hence $F$ has a unique minimizer $q_*\in(\epsilon,1)$, characterized by
\begin{equation}\label{eq:opt-q}
  \beta+\ln q_*=\frac{\pi(1-q_*)^2}{\sin(\pi q_*)} .
\end{equation}
Substituting \eqref{eq:opt-q} into the exponent of $F$ gives
\begin{equation}\label{eq:J-qstar}
  J(\epsilon)=F(q_*)=\cot\frac{\pi q_*}2\,(1-q_*)\,
  \exp\Bigl[\frac{\pi q_*(1-q_*)}{\sin(\pi q_*)}\Bigr].
\end{equation}

\paragraph{Asymptotics.}
As $\epsilon\to0$, we have $q_*\to0$. Indeed, for each fixed $q$, the left-hand side of
\eqref{eq:opt-q} tends to infinity, while the right-hand side remains unchanged.
For small $q$,
\[
  \frac{\pi(1-q)^2}{\sin(\pi q)}=\frac1q-2+O(q),\qquad
  \cot\frac{\pi q}2\,(1-q)\,\exp\Bigl[\frac{\pi q(1-q)}{\sin(\pi q)}\Bigr]
  =\frac{2\ee}{\pi}\Bigl(\frac1q-2+O(q)\Bigr).
\]
Thus \eqref{eq:opt-q} and \eqref{eq:J-qstar} become
\[
  \frac1{q_*}-2=\beta+\ln q_*+O(q_*),\qquad
  J(\epsilon)=\frac{2\ee}{\pi}\Bigl(\frac1{q_*}-2\Bigr)+O(q_*).
\]
The first relation implies $1/q_*=\beta-\ln\beta+2+o(1)$.
Substituting this into the second relation proves \eqref{eq:cost-asymptotics}.


\section{Discussion}
Our work establishes a worst-case lower bound for the LCHS query cost that matches the cost achievable with the kernel family in \cite{NY26}.
For any budget $(R,\lambda)$, our result determines the smallest achievable error $\epsilon$. Given a target error $\epsilon$ and a resource estimate $\Lambda(R,\lambda)$ that depends on $R$ and $\lambda$, we can optimize $\Lambda(R,\lambda)$ over the budgets admissible for that $\epsilon$. \Cref{cor:cost} treats the choice $\Lambda = \lambda R$ by solving the optimization problem \eqref{eq:opt prob}. The same approach applies to other cost models, such as
$\Lambda = \lambda \cdot \mathrm{HS}(R, \delta)$
where $\mathrm{HS}(R, \delta)$ is a uniform upper bound on the
query cost of implementing the controlled Hamiltonian evolutions
for $|k|\le R$ to accuracy $\delta$ using the chosen Hamiltonian simulation method. In this setting, the corresponding optimization problem is
$$\min_{\lambda, R, \delta}\  \lambda \cdot \mathrm{HS}(R, \delta)\quad \text{s.t.}\quad V_R(\lambda) + \lambda\delta\le \epsilon.$$

Our theorem establishes a worst-case lower bound over all dissipative evolutions. Improvements may still be possible for special classes of matrices, for example when the commutator $[X,Y]$ can be controlled or the matrix dimension $d$ is relatively small.

\section*{Acknowledgements}
The authors are supported by the U.S. Department of Energy, Office of Science, Accelerated Research in Quantum Computing Centers, Quantum Utility through Advanced Computational Quantum Algorithms, Grant No. DE-SC0025572. 

\bibliographystyle{amsalpha}
\bibliography{ref}

\appendix

\section{Technical lemmas}\label{app:technical}

\begin{lemma}[Contraction]\label{lem:contraction}
Let $X\succeq0$ and $Y=Y^*$, and let $\operatorname{Im}z\le0$.
Then $\|\ee^{-\ii t(zX+Y)}\|\le1$ for all $t\ge0$.
\end{lemma}

\begin{proof}
Let $G=-\ii(zX+Y)$ and $\vec{\xi}(t)=\ee^{tG}\vec{\xi}$.
Differentiating gives $\frac{\dd}{\dd t}\|\vec{\xi}(t)\|^2=2\operatorname{Re}\ip{\vec{\xi}(t)}{G\vec{\xi}(t)}$.
Since $\ip{\vec{\xi}}{X\vec{\xi}}\ge0$, $\ip{\vec{\xi}}{Y\vec{\xi}}\in\R$ and
$\operatorname{Re}(-\ii z)=\operatorname{Im}z$, it follows that
$\operatorname{Re}\ip{\vec{\xi}}{G\vec{\xi}}=\operatorname{Im}z\,\ip{\vec{\xi}}{X\vec{\xi}}\le0$.
Thus $\|\vec{\xi}(t)\|$ is nonincreasing.
\end{proof}

\begin{lemma}[Duhamel estimate]\label{lem:duhamel}
Let $G\in\C^{d\times d}$ satisfy $\|\ee^{-\ii tG}\|\le1$ for $t\ge0$.
Let $\vec{\psi}\in C^1([0,\tau];\C^d)$ satisfy
\[
  \ii\,\vec{\psi}'(t)=G\vec{\psi}(t)+\vec{r}(t),\qquad \|\vec{r}(t)\|\le\eta\quad(0\le t\le\tau),
\]
with $\vec{r}$ continuous.
Then $\|\vec{\psi}(\tau)-\ee^{-\ii\tau G}\vec{\psi}(0)\|\le\tau\eta$.
\end{lemma}

\begin{proof}
Let $\vec{F}(t)=\ee^{-\ii(\tau-t)G}\vec{\psi}(t)$.
Then
\[
  \vec{F}'(t)=\ee^{-\ii(\tau-t)G}\bigl(\ii G\vec{\psi}(t)+\vec{\psi}'(t)\bigr)
       =-\ii\,\ee^{-\ii(\tau-t)G}\vec{r}(t),
\]
so $\|\vec{F}'(t)\|\le\eta$.
Hence $\|\vec{\psi}(\tau)-\ee^{-\ii\tau G}\vec{\psi}(0)\|=\|\vec{F}(\tau)-\vec{F}(0)\|\le\tau\eta$.
\end{proof}

\end{document}